\documentclass[a4paper]{jpconf}
\usepackage{amsmath}
\usepackage{graphicx}

\newcommand{\KV}{{\mbox{$\kappa \sigma^{2}$}}}
\newcommand{\SD}{{\mbox{$S \sigma$}}}

\newcommand{ \be }{\begin{equation}}
\newcommand{ \ee }{\end{equation}}

\begin{document}
\title{Error Estimation for Moments Analysis in Heavy Ion Collision Experiment}

\author{Xiaofeng Luo}
\address{Key Laboratory of Quark and Lepton Physics (Huazhong Normal University), Ministry of Education}
\address{Institute of Particle Physics, Huazhong Normal University, Wuhan 430079, China}

\ead{xfluo@iopp.ccnu.edu.cn}

\begin{abstract}
Higher moments of conserved quantities are predicted to be sensitive
to the correlation length and connected to the thermodynamic
susceptibility. Thus, higher moments of net-baryon, net-charge and
net-strangeness have been extensively studied theoretically and
experimentally to explore phase structure and bulk properties of QCD
matters created in heavy ion collision experiment. As the higher
moments analysis is statistics hungry study, the error estimation is
crucial to extract physics information from the limited experimental
data. In this paper, we will derive the limit distributions and
error formula based on Delta theorem in statistics for various order
moments used in the experimental data analysis. The Monte Carlo
simulation is also applied to test the error formula.

\end{abstract}

\section{Introduction}
The main goal of performing the higher moments analysis is to study
the bulk properties, such as QCD phase transition~\cite{chiral_HRG},
QCD critical point~\cite{PRL,WWND2011,qcp_signal,ratioCumulant} and
thermalization~\cite{WWND2011}, of QCD matters created in the heavy
ion collisions experiment. It opens a completely new domain and
provides quantitative method for probing the bulk properties of the
hot dense nuclear  matter. On the other hand, the higher moment
analysis can be also used to constrain some fundamental parameters,
such as the scale for the QCD phase diagram (the transition
temperature $T_c$ at $\mu_B=0$), by comparing the experimental data
with the first principle Lattice QCD calculations~\cite{science}.

Higher moments (Variance ($\sigma^{2}$), Skewness ($S$), Kurtosis
($\kappa$) {\it etc.}) of conserved quantities, such as net-baryon,
net-charge, and net-strangeness, distributions can be directly
connected to the corresponding thermodynamic susceptibilities in
Lattice QCD~\cite{Lattice,MCheng2009} and Hadron Resonance Gas (HRG)
model~\cite{HRG}, for {\em e.g.} the third order susceptibility of
baryon number ($\chi^{(3)}_{B}$) is related to the third cumulant
($<(\delta N_{B})^{3}>$) of baryon number distributions as
$\chi^{(3)}_{B}={<(\delta N_{B})^{3}>}/{VT^3}$; $V,T$ are volume and
temperature of system respectively. As the volume of the system is
hard to determine, the susceptibility ratio, such as
$\chi^{(4)}_{B}$/$\chi^{(2)}_{B}$ and
$\chi^{(3)}_{B}$/$\chi^{(2)}_{B}$, are used to compare with the
experimental data as $\kappa
\sigma^2=\chi^{(4)}_{B}$/$\chi^{(2)}_{B}$ and $S
\sigma=\chi^{(3)}_{B}$/$\chi^{(2)}_{B}$. We also measure the ratios
of the sixth and eighth to second order cumulants of the net-baryon
number fluctuations, as $\chi^{(6)}_{B}$/$\chi^{(2)}_{B}$ and
$\chi^{(8)}_{B}$/$\chi^{(2)}_{B}$, respectively, which are predicted
to be with negative value when the freeze out temperature is close
to the the chiral phase transition temperature. Theoretical
calculations demonstrate that the experimental measurable net-proton
(proton number minus anti-proton number) number fluctuations can
effectively reflect the fluctuations of the net-baryon
number~\cite{Hatta}. Thus, it is of great interest to measure the
higher moments of event-by-event net-proton multiplicity
distributions in the heavy ion collision experiment.

In section 2, we will show the definition of central moments and
cumulants. Then, the Delta theorem in statistics will be discussed
in section 3 and applied to derive the error formula for various
order moments. In section 4, Monte Carlo simulation has been done to
check the validity of the error formula. The summary and conclusion
will go to the chapter 5.

\section{Central Moments and Cumulants of Event-by-Event Fluctuations} In
statistics, probability distribution functions can be characterized
by the various moments, such as mean ($M$), variance ($\sigma^2$),
skewness ($S$) and kurtosis ($\kappa$). Before introducing the above
moments used in our analysis, we would like to define central
moments and cumulants, which are alternative methods to describe a
distribution.

Experimentally, we measure net-proton number event-by-event wise,
$N_{p-\bar{p}}=N_{p}-N_{\bar{p}}$, which is proton number minus
antiproton number. In the following, we use $N$ to represent the
net-proton number $N_{p-\bar{p}}$ in one event. The average value
over whole event ensemble is denoted by $\hat \mu=<N>$, where the
single angle brackets are used to indicate ensemble average of an
event-by-event distributions and the hat symbol denotes the sample
estimator.

The deviation of $N$ from its mean value are defined by
\begin{equation}
  \delta N=N-<N>=N-\hat \mu.
\end{equation}

The $r^{th}$ order sample estimates for central moments are defined
as:

\begin{eqnarray}
 \hat \mu _r  &=&  < (\delta N)^r  >  \\
 \hat \mu _1  &=& 0
\end{eqnarray}

Then, we can define the sample estimates for various order cumulants
of event-by-event distributions as:
\begin{eqnarray}
 \hat C_1  &=& \hat \mu \\
 \hat C_2 & = &\hat \mu _2  \\
 \hat C_3  &=& \hat \mu _3  \\
\hat C_n(n>3)  &=& \hat \mu _n  - \sum\limits_{m = 2}^{n - 2}
{\left(
\begin{array}{l}
 n - 1 \\
 m - 1 \\
 \end{array} \right)\hat C_m } \hat \mu _{n - m}
\end{eqnarray}

An important property of the cumulants is their additivity for
independent variables. If $X$ and $Y$ are two independent random
variables, then we have $C_{i,X+Y}=C_{i,X}+C_{i,Y}$ for $i$th order
cumulant.

Once we have the definition of cumulants, sample estimators for
skewness and kurtosis can be denoted as:
\begin{eqnarray}
\hat M=\hat C_{1,N},\hat \sigma^{2}=\hat C_{2,N},\hat S=\frac{\hat
C_{3,N}}{(\hat C_{2,N})^{3/2}},\hat \kappa=\frac{\hat C_{4,N}}{(\hat
C_{2,N})^{2}}
\end{eqnarray}

Then, the moments product $\hat  \kappa \hat \sigma^{2}$ and $ \hat
S \hat \sigma$ can be expressed in term of cumulant ratio.

\begin{eqnarray}
\hat \kappa \hat \sigma^{2}=\frac{\hat C_{4,N}}{\hat C_{2,N}}, \hat
S\hat \sigma=\frac{\hat C_{3,N}}{\hat C_{2,N}}.
\end{eqnarray}

With above definition of various moments, we can calculate various
moments and moment products with the measured event-by-event
net-proton distributions.

\section{Delta Theorem in Statistics}
Before deriving the limit distributions as well as the error formula
of various moments and moment products, we would like to introduce
you the delta theorem that used in the calculations. The delta
theorem~\cite{statistics1,statistics2,statistics3} says how to
approximate the distribution of a transformation of a statistic in
large samples if we can approximate the distribution of the
statistic itself. Distributions of transformations of a statistic
are of great importance in applications. We will give the theorem
with one and multi-dimensional cases without proofs. Before
introducing the delta theorem, we will show you an useful theorem of
sample moments~\cite{statistics2}.

{\bf {\textit {Theorem A}}:} If central moments $\mu _{2k}  = E[(X -
\mu )^{2k}] <\infty$, then the random vector $\sqrt n (\hat \mu _2 -
\mu _2 ,...\hat \mu _k  - \mu _k )$ converges in distributions to
$(k-1)$-variate normal with mean vector $(0,0,0,...,0)$ and
covariance matrix $[\Sigma _{ij} ]_{(k - 1) \times (k - 1)} $, where
$$\Sigma _{ij}  = \mu _{i + j}  - \mu _i \mu _j  - i\mu _{i - 1} \mu
_{j + 1}  - j\mu _{i + 1} \mu _{j - 1}  + ij\mu _{i - 1} \mu _{j -
1} \mu _2$$ For instance, we have the limit distribution for the
sample variance $\mu_2=\hat \sigma^2$:
$$\hat \sigma ^2\xrightarrow[]{d} N(\sigma ^2 ,\frac{{\mu _4  - \sigma ^4 }}{n})
$$ Then the variance of the sample variance is $Var(\hat \sigma^2)=({\mu_4 -\sigma^4
})/{n}$.

In the following, we will introduce the one and multi-dimension
delta theorems and their applications.

{ {\bf { \textit {Delta Theorem-I (One-dimension)}:}}} Suppose that
random variable $X$ distribute as $N(\mu,\frac{{\sigma^2}}{n})$, let
$g$ be a real-valued function differentiable at $x=\mu$, with
$g^{'}(\mu ) \ne0$. Then we get the limit distribution of $g(X)$:
$$g(X) \xrightarrow[]{d} N(g(\mu ),[g^{'} (\mu )]^2 \frac{{\sigma ^2
}} {n})$$

As an application of the delta theorem, let's estimate the limit
distribution of the sample standard deviation $\hat \sigma$. It was
seen in theorem A that $$\hat \sigma ^2\xrightarrow[]{d} N(\sigma ^2
,\frac{{\mu _4 - \sigma ^4 }}{n})$$ It follows that the sample
standard deviation is also asymptotically normal, namely $$ \hat
\sigma \xrightarrow[]{d} N(\sigma  ,\frac{{\mu _4  - \sigma ^4
}}{{4\sigma ^2 n}})$$, with the function $g(x)=\sqrt{x}$.

The following theorem extends the above delta theorem to the case of
a vector-valued function $g$ to a vector ${\bf{
X}}=\{X_{1},X_{2},...,X_{k}\}$.

{ \bf {\textit {Delta Theorem-II (Multi-dimension)}}:} Suppose that
${\bf{ X}}=\{X_{1},X_{2},...,X_{k}\}$ is normally distributed as
$N({\bf{\mu}}, {\bf{\Sigma}}/n)$,  with $\bf \Sigma$ a covariance
matrix. Let ${\bf {g(x)}}=( g_{1}(x),...,g_{m}(x))$, ${\bf
{x}}=(x_{1},...x_{k})$, be a vector-valued function for which each
component function $g_{i}(x)$ is real-valued and has a non-zero
differential $g_{i}(\mu)$, at ${\bf{x}}={\bf{\mu}}$. Put $${\bf{D}}
= \left[ {\left. {\frac{{\partial g_i }}{{\partial x_j }}}
\right|_{x = \mu } } \right]_{m \times k}$$ Then
$${\bf{g}}({\bf{X}})\xrightarrow[]{d}N({\bf{g}}(\mu),\frac{{\bf{D\Sigma D^{'}}}}{n})$$

In the following sub-sections, we will derive the joint limiting
distributions for  higher order moments ($\sigma, S, \kappa$) and
moment products ($S\sigma, \kappa \sigma^2, \kappa \sigma/S$). The
limit distributions of sixth and eighth to second order cumulants
ratio will be also calculated.

\subsection{\bf Joint Limiting Distributions of Sample Standard Deviation($\hat \sigma$), Skewness ($\hat S$) and Kurtosis($\hat \kappa$)}

The multi-dimension delta theorem will be applied to derive the
joint limit distributions for the sample statistic vector $$
{\bf{\hat T}} = \left( \begin{array}{l}
 {\hat \sigma } \\
 {\hat S} \\
 {\hat \kappa } \\
 \end{array} \right)$$
For the sample moments vector $${\bf{\hat W}} = \left(
\begin{array}{l}
 \hat \mu _2  \\
 \hat \mu _3  \\
 \hat \mu _4  \\
 \end{array} \right)$$
We have the limit distributions, when the sample is large enough:
$$ {\bf{\hat W}} = \left(
\begin{array}{l}
 \hat \mu _2  \\
 \hat \mu _3  \\
 \hat \mu _4  \\
 \end{array} \right)\xrightarrow[]{d}N(\left( \begin{array}{l}
 \mu _2  \\
 \mu _3  \\
 \mu _4  \\
 \end{array} \right),\frac{{\bf{\Sigma}} }{n})$$, where the
 ${\bf{\Sigma}}$ is the $3\times3$ covariance matrix of the multi-variate
 vector ${\bf{\hat W}}$. The covariance matrix is a symmetrical
 matrix and the matrix element can be calculated via theorem A.
 $$
\begin{array}{l}
 \Sigma _{11}  = Var(\hat \mu _2 ) = \mu _4  - \sigma ^4  \\
 \Sigma _{22}  = Var(\hat \mu _3 ) = \mu _6  - \mu _3 ^2  - 6\mu _4 \sigma ^2  + 9\sigma ^6  \\
 \Sigma _{33}  = Var(\hat \mu _4 ) = \mu _8  - \mu _4 ^2  - 8\mu _3 \mu _5  + 16\mu _3 ^2 \sigma ^2  \\
 \Sigma _{12}  = \Sigma _{21}  = Cov(\hat \mu _2 ,\hat \mu _3 ) = \mu _5  - 4\mu _3 \sigma ^2  \\
 \Sigma _{13}  = \Sigma _{31}  = Cov(\hat \mu _2 ,\hat \mu _4 ) = \mu _6  - 4\mu _3 ^2  - \mu _4 \sigma ^2  \\
 \Sigma _{23}  = \Sigma _{32}  = Cov(\hat \mu _3 ,\hat \mu _4 ) = \mu _7  - 3\mu _5 \sigma ^2  - 5\mu _3 \mu _4  + 12\mu _3 \sigma ^4  \\
 \end{array}
$$
Based on the definition of the Standard deviation, Skewness and
Kurtosis, we define a function vector ${\bf{g}}=(g_1=\sqrt{
\mu_2},g_2= \mu_3/( \mu_2)^{3/2},g_3=\mu_4/( \mu_2)^{2}-3)$. Then we
have:
$${\bf{D}} = \left. {\left[ {\frac{{\partial g_i }}{{\partial \mu _j }}}
\right]} \right|_{3 \times 3}  = \left( {\begin{array}{*{20}c}
   {\frac{{\partial g_1 }}{{\partial \mu _2 }}} & {\frac{{\partial g_1 }}{{\partial \mu _3 }}} & {\frac{{\partial g_1 }}{{\partial \mu _4 }}}  \\
   {\frac{{\partial g_2 }}{{\partial \mu _2 }}} & {\frac{{\partial g_2 }}{{\partial \mu _3 }}} & {\frac{{\partial g_2 }}{{\partial \mu _4}}}  \\
   {\frac{{\partial g_3 }}{{\partial \mu _2 }}} & {\frac{{\partial g_3 }}{{\partial \mu _3 }}} & {\frac{{\partial g_3 }}{{\partial \mu _4 }}}  \\
\end{array}} \right) =\left( {\begin{array}{*{20}c}
   {1/(2\sigma )} & 0 & 0  \\
   { - 3\mu _3 /(2\sigma ^5) } & {1/\sigma ^3 } & 0  \\
   { - 2\mu _4 /\sigma ^6 } & 0 & {1/\sigma ^4 }  \\
\end{array}} \right)$$
Then, according to the multi-variate delta theorem, we have the
joint limiting distribution for the random sample vector ${\bf{\hat
T}}$
$${\bf{\hat T}}\xrightarrow[]{d}N(\left( \begin{array}{l}
 \sigma  \\
 S \\
 \kappa  \\
 \end{array} \right),\frac{{{\bf{D\Sigma D}}^{\bf{'}} }}{n})$$
, where the covariance matrix $ {\bf{\Gamma  = D\Sigma
D}}^{\bf{'}}/n $ is a $3\times3$ symmetrical matrix. The matrix
element:$$
\begin{array}{l}
 \Gamma _{{\rm{11}}} {\rm{ = }}Var(\hat \sigma ) = (m_4  - 1)\sigma ^2 /(4n) \\
 \Gamma _{22}  = Var(\hat S) = [9 - 6m_4  + m_3 ^2 (35 + 9m_4 )/4 - 3m_3 m_5  + m_6 ]/n \\
 \Gamma _{33}  = Var(\hat \kappa ) = [- m_4 ^2  + 4m_4 ^3  + 16m_3 ^2 (1 + m_4 ) - 8m_3 m_5  - 4m_4 m_6  + m_8 ]/n \\
 \Gamma _{{\rm{12}}}  = \Gamma _{{\rm{21}}}  = Cov(\hat \sigma ,\hat S) =  - [m_3 {\rm{ }}(5{\rm{ }} + {\rm{ }}3{\rm{ }}m_4 ){\rm{ }} - {\rm{ }}2{\rm{ }}m_5 {\rm{] }}\sigma /(4n) \\
 \Gamma _{13}  = \Gamma _{31}  = Cov(\hat \sigma ,\hat \kappa ){\rm{ = [}}( - 4m_3 ^2  + m_4  - 2m_4 ^2  + m_6 )\sigma ]/(2n) \\
 \Gamma _{23}  = \Gamma _{33}  = Cov(\hat S,\hat \kappa ) = [6m_3 ^3  - (3 + 2m_4 )m_5  + 3m_3 (8 + m_4  + 2m_4 ^2  - m_6 )/2 + m_7 ]/n \\
 \end{array}
$$, where the normalized central moments $m_{r}=\mu_{r}/\sigma^{r}$.
The non-zero values for the non-diagonal elements of the covariance
matrix indicate there are correlation between those three moments
$\sigma, S, \kappa$. When the distribution is a symmetrical
distribution, the odd normalized central moments will be zero, thus
we have $ Cov(\hat \sigma ,\hat S)= Cov(\hat S,\hat \kappa)=0$,
which means there has no correlation between skewness and the other
two moments. For normal distributions,

$$\begin{array}{l}
 \Gamma _{{\rm{11}}} {\rm{ = }}Var(\hat \sigma ) = \sigma ^2 /(2n) \\
 \Gamma _{22}  = Var(\hat S) = 6/n \\
 \Gamma _{33}  = Var(\hat \kappa ) = 24/n \\
 \Gamma _{{\rm{12}}}  = \Gamma _{{\rm{21}}}  = Cov(\hat \sigma ,\hat S) = 0 \\
 \Gamma _{13}  = \Gamma _{31}  = Cov(\hat \sigma ,\hat \kappa ){\rm{ = }}0 \\
 \Gamma _{23}  = \Gamma _{32}  = Cov(\hat S,\hat \kappa ) = 0 \\
 \end{array}
$$ The non-diagonal matrix elements are zero and no correlations
between $\sigma, S, \kappa$.

\subsection{\bf Joint Limiting Distributions of Sample Moment Products ($\hat S \hat \sigma$,  $\hat \kappa\hat \sigma^2$,$\hat \kappa \hat \sigma/\hat S$)}
To derive the joint limiting distributions for the moment products
($\hat S \hat \sigma$,  $\hat \kappa\hat \sigma^2$,$\hat \kappa \hat
\sigma/\hat S$), we apply similar procedures as above. Define the
random vector: $$ {\bf{\hat T}} = \left( \begin{array}{l}
 \hat S\hat \sigma  \\
 \hat \kappa \hat \sigma ^2  \\
 \frac{{\hat \kappa \hat \sigma }}{{\hat S}} \\
 \end{array} \right)$$ Based on the definition of the moment products, we define a function vector ${\bf{g}}=(
g_1  = \hat \mu _3 /\hat \mu _2 ,g_2  = \hat \mu _4 /\hat \mu _2  -
3\hat \mu _2 ,g_3  = (\hat \mu _4  - 3\hat \mu _2 ^2 )/\hat \mu _3
)$. Then we have:
$${\bf{D}} = \left. {\left[ {\frac{{\partial g_i }}{{\partial \mu _j }}}
\right]} \right|_{3 \times 3}  = \left( {\begin{array}{*{20}c}
   {\frac{{\partial g_1 }}{{\partial \mu _2 }}} & {\frac{{\partial g_1 }}{{\partial \mu _3 }}} & {\frac{{\partial g_1 }}{{\partial \mu _4 }}}  \\
   {\frac{{\partial g_2 }}{{\partial \mu _2 }}} & {\frac{{\partial g_2 }}{{\partial \mu _3 }}} & {\frac{{\partial g_2 }}{{\partial \mu _4 }}}  \\
   {\frac{{\partial g_3 }}{{\partial \mu _2 }}} & {\frac{{\partial g_3 }}{{\partial \mu _3 }}} & {\frac{{\partial g_3 }}{{\partial \mu _4 }}}  \\
\end{array}} \right) = \left( {\begin{array}{*{20}c}
   { - \mu _3 /\sigma ^4 } & {1/\sigma ^2 } & 0  \\
   { { - (\mu _4 /\sigma ^4  + 3)}} & 0 & {1/\sigma ^2 }  \\
   { - 6\sigma ^2 /\mu _3 } & { - (\mu _4  - 3\sigma ^4 )/\mu _3 ^2 } & {1/\mu _3 }  \\
\end{array}} \right)
$$ Then, according to the multi-variate delta theorem, we have the
joint limiting distribution for the random sample vector ${\bf{\hat
T}}$ $${\bf{\hat T}}\xrightarrow[]{d}N(\left( \begin{array}{l}
 S \sigma  \\
 \kappa \sigma^{2} \\
 \kappa \sigma/S \\
 \end{array} \right),\frac{{{\bf{D\Sigma D}}^{\bf{'}} }}{n})$$ The
corresponding matrix element of covariance matrix ${\bf{\Pi  =
D\Sigma D}}^{\bf{'}}/n $ are \begin{eqnarray} \nonumber
\Pi_{{\rm{11}}}&=&Var(\hat S\hat \sigma )=[9 - 6m_4  + m_3 ^2 (6 +
m_4 ) - 2m_3 m_5  + m_6 ]\sigma ^2 /n \\ \nonumber \Pi _{22} &=&
Var(\hat \kappa \hat \sigma ^2 ) = [ - 9 + 6m_4 ^2  + m_4 ^3  + 8m_3
^2 (5 + m_4 ) - 8m_3 m_5  + m_4 (9 - 2m_6 ) - 6m_6  + m_8 ]\sigma ^4
/n \\ \nonumber \Pi_{33}  &=& Var(\hat \kappa \hat \sigma /\hat S) =
[64m_3 ^4  - 8m_3 ^3 m_5  - ( - 3 + m_4 )^2 ( - 9 + 6m_4 - m_6 ) +
2m_3 ( - 3 + m_4 )(9m_5  - m_7 ) \\ \nonumber
  &+& m_3 ^2 (171 - 48m_4  + 8m_4 ^2  - 12m_6  + m_8 )]\sigma ^2 /(n \times m_3 ^4 )
  \\ \nonumber
 \Pi _{{\rm{12}}}  &=& \Pi _{{\rm{21}}}  = Cov(\hat S\hat \sigma ,\hat \kappa \hat \sigma ^2 )= [4m_3 ^3  - (6 + m_4 )m_5  + m_3 (21 + 2m_4  + m_4 ^2  - m_6 ) + m_7 ]\sigma ^3 /n
 \\ \nonumber
 \Pi _{13}  &=& \Pi _{31}  = Cov(\hat S\hat \sigma ,\hat \kappa \hat \sigma /\hat S)=[4m_3 ^4  + ( - 3 + m_4 )( - 9 + 6m_4  - m_6 ) - m_3 ^2 ( - 39 + m_4  + m_6 )
 \\ \nonumber
  &+& m_3 (( - 12 + m_4 )m_5  + m_7 )]\sigma ^2 /(n \times m_3 ^2 )
  \\ \nonumber
 \Pi _{23}  &=& \Pi _{32}  = Cov(\hat \kappa \hat \sigma ^2 ,\hat \kappa \hat \sigma /\hat S) = [4m_3 ^3 (13 + m_4 ) - 8m_3 ^2 m_5  + ( - 3 + m_4 )((6 + m_4 )m_5  - m_7 )
 \\ \nonumber
  &+& m_3 (54 + 7m_4 ^2  - 9m_6  - m4(6 + m_6 ) + m_8 )]\sigma ^3 /(n \times m_3 ^2 )
  \\ \nonumber \end{eqnarray}, where the normalized central moments
 $m_{r}=\mu_{r}/\sigma^{r}$. Supposing that the distribution is
 symmetrically distributed, the variance of the sample statistic $\hat \kappa \hat \sigma/\hat S$
and the corresponding covariance are not well defined. For normal
distribution, we have $$
\begin{array}{l}
 \Pi _{{\rm{11}}} {\rm{ = }}Var(\hat S\hat \sigma ) = 6\sigma ^2 /n \\
 \Pi _{22}  = Var(\hat \kappa \hat \sigma ^2 ) = 24\sigma ^4 /n \\
\Pi _{{\rm{12}}}  = \Pi _{{\rm{21}}}  = Cov(\hat S\hat \sigma ,\hat
\kappa \hat \sigma ^2 ) = 0  \end{array}$$ there has no correlation
between sample statistic $\hat S \hat\sigma$ and $\hat \kappa \hat
\sigma^{2}$.
\subsection{\bf Limit Distribution for Sample Cumulant ratio ($\hat C_{6}/\hat C_{2}$ and $\hat C_{8}/\hat C_{2}$ )}
We only derive the limit distribution for sixth to second cumulant
order ratio and give the results for $\hat C_{8}/\hat C_{2}$ without
derivation. For the sample moments vector $${\bf{\hat W}} = \left(
\begin{array}{l}
 \hat \mu _2  \\
 \hat \mu _3  \\
 \hat \mu _4  \\
 \hat \mu_6 \\
  \end{array} \right)$$
We have the limit distributions, when the sample is large enough:
$$ {\bf{\hat W}} = \left(
\begin{array}{l}
 \hat \mu _2  \\
 \hat \mu _3  \\
 \hat \mu _4  \\
 \hat \mu_6 \\
 \end{array} \right)\xrightarrow[]{d}N(\left( \begin{array}{l}
 \mu _2  \\
 \mu _3  \\
 \mu _4  \\
 \mu_6    \\
 \end{array} \right),\frac{{\bf{\Omega}} }{n})$$, where the
 ${\bf{\Omega}}$ is the $4\times4$ covariance matrix of the multi-variate
 vector ${\bf{\hat W}}$. The covariance matrix is a symmetrical
 matrix and the matrix element can be calculated via theorem A.
 $$\begin{array}{l}
 \Omega _{11}  = Var(\hat \mu _2 ) = \mu _4  - \sigma ^4  \\
 \Omega _{22}  = Var(\hat \mu _3 ) = \mu _6  - \mu _3 ^2  - 6\mu _4 \sigma ^2  + 9\sigma ^6  \\
 \Omega _{33}  = Var(\hat \mu _4 ) = \mu _8  - \mu _4 ^2  - 8\mu _3 \mu _5  + 16\mu _3 ^2 \sigma ^2  \\
 \Omega _{44}  = Var(\hat \mu _6 ) = \mu _{12}  - 12\mu _5 \mu _7  - \mu _6 ^2  - 36\mu _5 ^2 \sigma ^2  \\
 \Omega _{12}  = \Omega _{21}  = Cov(\hat \mu _2 ,\hat \mu _3 ) = \mu _5  - 4\mu _3 \sigma ^2  \\
 \Omega _{13}  = \Omega _{31}  = Cov(\hat \mu _2 ,\hat \mu _4 ) = \mu _6  - 4\mu _3 ^2  - \mu _4 \sigma ^2  \\
 \Omega _{14}  = \Omega _{41}  = Cov(\hat \mu _2 ,\hat \mu _6 ) = \mu _8  - \mu _6 \sigma ^2  - 6\mu _3 \mu _5  \\
 \Omega _{23}  = \Omega _{32}  = Cov(\hat \mu _3 ,\hat \mu _4 ) = \mu _7  - 3\mu _5 \sigma ^2  - 5\mu _3 \mu _4  + 12\mu _3 \sigma ^4  \\
 \Omega _{24}  = \Omega _{42}  = Cov(\hat \mu _3 ,\hat \mu _6 ) = \mu _9  - 3\mu _7 \sigma ^2  - \mu _3 \mu _6  - 6\mu _4 \mu _5  + 18\mu _5 \sigma ^4  \\
 \Omega _{34}  = \Omega _{43}  = Cov(\hat \mu _4 ,\hat \mu _6 ) = \mu _{10}  - 4\mu _3 \mu _7  - \mu _4 \mu _6  - 6\mu _5 ^2  + 24\mu _3 \mu _5 \sigma ^2  \\
 \end{array}$$ We define a function $g(\mu _2 ,\mu _3 ,\mu _4 ,\mu _6 ) = (\mu _6  - 15\mu _2 \mu _4  -
10\mu _3 ^2  + 30\mu _2 ^3 )/\mu _2 $, then we have the gradient
matrix:
$$
{\bf{D}} = \left. {\left[ {\frac{{\partial g}}{{\partial \mu _j }}}
\right]} \right|_{1 \times 4}  = \{ ( - \frac{{\mu _6 }}{{\mu _2 ^2
}} + 10\frac{{\mu _3 ^2 }}{{\mu _2 ^2 }} + 60\mu _2 ), -
20\frac{{\mu _3 }}{{\mu _2 }}, - 15,\frac{1}{{\mu _2 }}\}
$$ Following the delta theorem, we can obtain the limit
distribution of the sample cumulant ratio $\hat C_{6}/\hat C_{2}$.
$$\frac{{\hat C_6 }}{{\hat C_2 }} \xrightarrow[]{d} N(\frac{{C_6 }}{{C_2
}},\frac{{{\bf{D\Omega D}}^{\bf{'}} }}{n})$$ Then, the variance of
the sample statistic:
\begin{eqnarray}
 \nonumber
 Var(\frac{{\hat C_6 }}{{\hat C_2 }}) = \frac{{{\bf{D\Omega D}}^{\bf{'}} }}{n} &= &[10575 - 30m_{10}  + m_{12}  + 18300m_3 ^2  + 2600m_3 ^4  - 225( - 3 + m_4 )^2  - 7440m_3 m_5
 \\ \nonumber
  &- &520m_3 ^3 m_5  + 216m_5 ^2  - 2160m_6  - 200m_3 ^2 m_6  + 52m_3 m_5 m_6  + 33m_6 ^2
  \\   \nonumber
 & +& ( - 3 + m_4 )(10(405 - 390m_3 ^2  + 10m_3 ^4  + 24m_3 m_5 ) - 20(6 + m_3 ^2 )m_6  + m_6 ^2 )
 \\  \nonumber
  &+ &840m_3 m_7  - 12m_5 m_7  + 345m_8  + 20m_3 ^2 m_8  - 2m_6 m_8  - 40m_3 m_9 ]\sigma ^8/n
  \\  \nonumber
\end{eqnarray}, where the normalized central moments
 $m_{r}=\mu_{r}/\sigma^{r}$. For normal distribution, we have:$$Var(\frac{{\hat C_6 }}{{\hat C_2 }}) = \frac{{720\sigma ^8 }}{n}
$$

With similar procedure, we can obtain the variance of eighth to
second order cumulant ratio:

\begin{eqnarray}
\nonumber
 Var(\frac{{\hat C_8 }}{{\hat C_2 }}) &=& [198450 + 1204m_{12}  - 56m_{14}  + m_{16}  + 5376m_{11} m_3  - 112m_{13} m_3  - 9878400m_3 ^2
 \\ \nonumber
  &-& 1254400m_3 ^4  + 46550( - 3 + m_4 )^4  + 1225( - 3 + m_4 )^5  - 112m_{11} m_5  - 1270080m_3 m_5
  \\ \nonumber
  &-& 250880m_3 ^3 m_5 {\rm{ + }}176400m_5 ^2  + 169344m_3 ^2 m_5 ^2  - 6272m_3 m_5 ^3  - 114660m_6  + 1693440m_3 ^2 m_6
  \\ \nonumber
  &-& 142688m_3 m_5 m_6  + 3136m_5 ^2 m_6  - 784m_6 ^2  + 698880m_3 m_7  - 48384m_5 m_7  - 7168m_3 ^2 m_5 m_7
  \\ \nonumber
  &+& 896m_3 m_6 m_7  + 512m_7 ^2  + 63630m_8  - 118720m_3 ^2 m_8  - 112m_3 m_5 m_8  + 112m_5 ^2 m_8  - 420m_6 m_8
  \\ \nonumber
  &+& 128m_3 m_7 m_8  + 59m_8 ^2  - 2m_{10} ( - 7( - 885 + 224m_3 ^2  + 8m_3 m_5 ) + m_8 )
  \\ \nonumber
  &-& 70( - 3 + m_4 )^3 ( - 7(1125 + 80m_3 ^2  + 8m_3 m_5 ) + 70m_6  + m_8 )
  \\ \nonumber
  &+& 70( - 3 + m_4 )^2 (5040 + m_{10}  + 30240m_3 ^2  + 560m_3 m_5  - 56m_5 ^2  - 1078m_6  - 64m_3 m_7  + 21m_8 )
  \\ \nonumber
  &-& 92960m_3 m_9  + 3808m_5 m_9  - 16m_7 m_9  + ( - 3 + m_4 )( - 1488375 + 5180m_{10}  - 140m_{12}
  \\ \nonumber
  &-& 13524000m_3 ^2  + 4104240m_3 m_5  + 62720m_3 ^3 m_5  - 155232m_5 ^2  + 3136m_3 ^2 m_5 ^2
  \\ \nonumber
  &+& 343980m_6  - 62720m_3 ^2 m_6  - 7840m_3 m_5 m_6  - 295680m_3 m_7  + 9856m_5 m_7  - 67830m_8
  \\ \nonumber
  &-& 1120m_3 ^2 m_8  - 112m_3 m_5 m_8  + 140m_6 m_8  + m_8 ^2  + 8960m_3 m_9 )]\sigma ^{{\rm{12}}} /n
\end{eqnarray}
For normal distribution, we have:
$$Var(\frac{{\hat C_8 }}{{\hat C_2 }}) = \frac{{{\rm{40320}}\sigma
^{12} }}{n} $$

\section{Monto Carlo Simulation}
To check whether our results for the errors of the various moments
is reasonable or not, we have done Monte Carlo simulations.
Experimentally, we calculated the various moments from measured
event-by-event net-proton or net-charge multiplicity distributions.
For simplify, we assume the particle and anti-particle are
independently distributed as Poissonian distribution, which is an
appropriate approximation. Then, the difference of two independent
Poisson distributions distributed as "Skellam" distribution. Its
probability density distribution is
$$f(k;\mu _1 ,\mu _2 ) = e^{ -
(\mu _1 {{ + }}\mu _{{2}} )} (\frac{{\mu _1 }}{{\mu _{{2}} }})^{k/2}
I_{|k|} (2\sqrt {\mu _1 \mu _2 } ) $$, where the $\mu_1$ and $\mu_2$
are the mean value of two Poisson distributions, respectively, the
$I_{k}(z)$ is the modified bessel function of the first kind. Then,
we can calculate various moments ($M,\sigma,S,\kappa$) and moment
products ({\KV},{\SD}) products of the Skellam distribution. The
results are shown below:
\begin{eqnarray} \nonumber M&=&\mu_1-\mu_2 \\ \nonumber
\sigma&=&\sqrt{\mu_1+\mu_2} \\ \nonumber
S&=&\frac{\mu_1-\mu_2}{({\mu_1+\mu_2})^{3/2}} \\ \nonumber
\kappa&=&\frac{1}{\mu_1+\mu_2} \\ \nonumber
  S \sigma&=&\frac{\mu_1-\mu_2}{\mu_1+\mu_2} \\ \nonumber
  \kappa \sigma^{2}&=&\frac{C_{6}}{C_{2}}=\frac{C_{8}}{C_{2}}=1
\end{eqnarray}

\begin{figure}[htb]
\begin{center}
\includegraphics[width=0.85\textwidth]{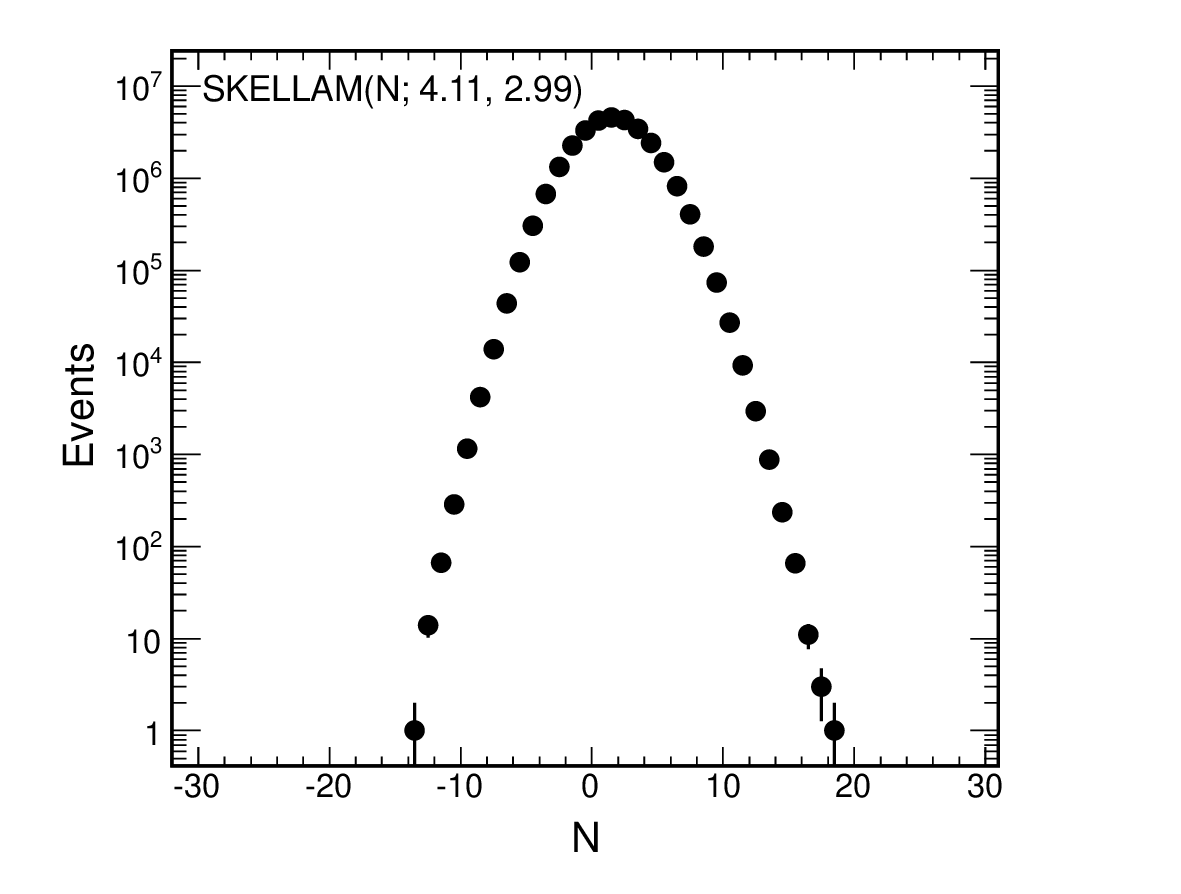}
\caption[]{Sample distribution (30 million events) with the
population distributed as skellam distribution.} \label{fig:skellam}
\end{center}\end{figure}

To do the simulation, we set the two mean values of the Skellam
distributions as $\mu_1=4.11,\mu_2=2.99$, which are similar with the
mean proton and anti-proton number in most central Au+Au collisions
at $\sqrt{s_{NN}}=200$ GeV measured by STAR experiment. Then, we can
generate random numbers as per the "Skellam" distribution. Fig.
\ref{fig:skellam} shows a distribution sampled from the "skellam"
population with 30 million events.

\begin{figure}[htb]
\begin{center}
\includegraphics[width=0.85\textwidth]{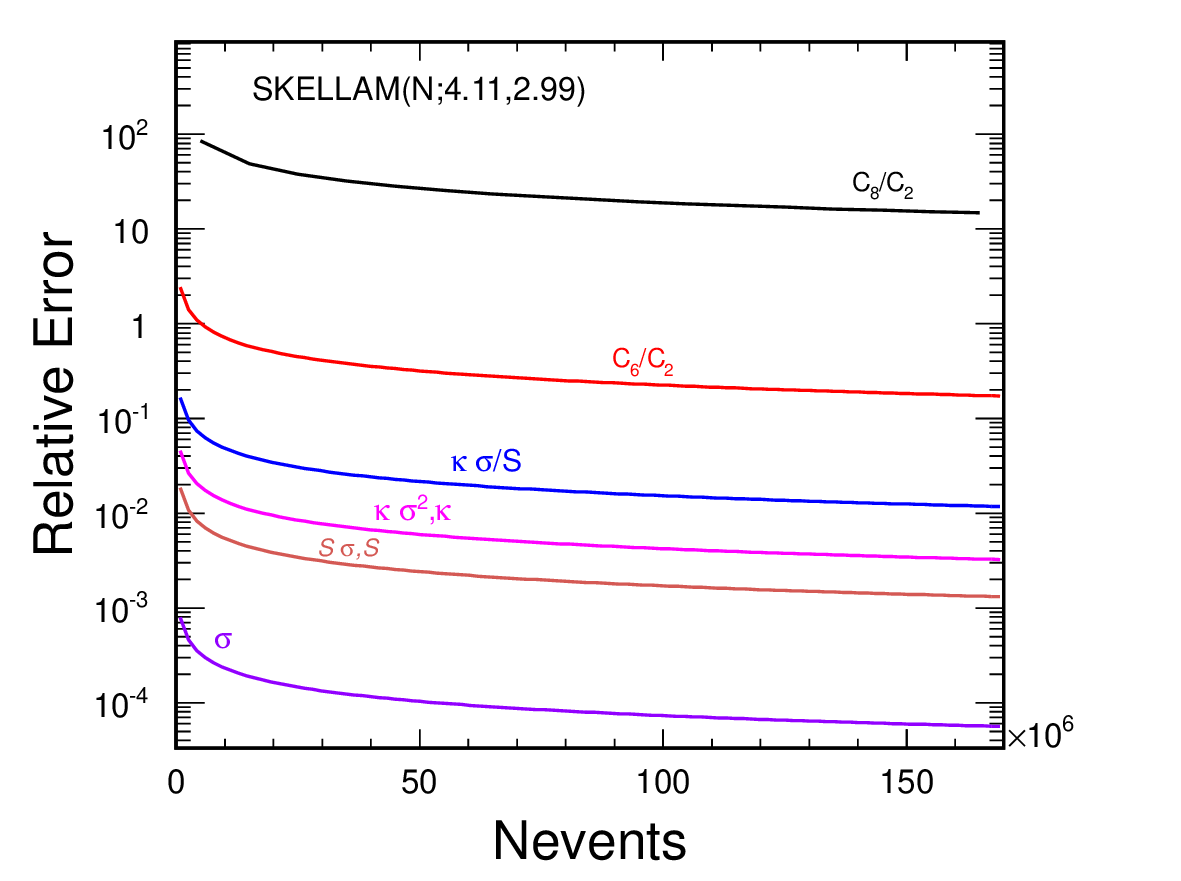}
\caption[]{Relative error as a function of number of events for various sample moments calculated from
the error formula shown in section 3 by assuming the population is with
"Skellam" distribution. } \label{fig:relative_error}
\end{center}\end{figure}

In Fig. \ref{fig:relative_error}, we show the relative error as a
function of events for various sample moments when the population is
with "Skellam" distribution. We may find that the higher order
moments are with larger relative errors, especially for the sixth
and eighth order to second cumulant ratios. As the input two
parameters for "Skellam" distribution $\mu_1,\mu_2$, are similar
with the mean proton and anti-proton number for Au+Au 200 GeV data,
we can estimate the number of events needed to achieve relative
small errors for the moments studied at this energy. In year 2010
and 2011, STAR experiment has accumulated few hundreds million
events of Au+Au collisions at $\sqrt{s_{NN}}$= 200 GeV both for
minbias and central trigger, from the Fig. \ref{fig:relative_error},
it allows us to study the higher moments of net-proton distributions
up to sixth order with the acceptable errors.

According to the error formula shown in Section 3,  for normal
distributions, the errors for cumulants ratios are proportional to
the standard deviation of the distribution as£º
$$\begin{array}{l}
 error(\hat S\hat \sigma ) \propto \frac{\sigma }{{\sqrt n }} \\
 error(\hat \kappa \hat \sigma ^2 ) \propto \frac{{\sigma ^2 }}{{\sqrt n }} \\
 error(\frac{{\hat C_6 }}{{\hat C_2 }}) \propto \frac{{\sigma ^4 }}{{\sqrt n }} \\
 error(\frac{{\hat C_8 }}{{\hat C_2 }}) \propto \frac{{\sigma ^6 }}{{\sqrt n }} \\
 \end{array}$$

Thus, with similar phase space coverage and number of events as high
collision energy, we may get larger errors when we are doing moments
analysis of net-proton distributions at low energy, as more nucleons
are expected to be stopped in the central region due to nuclear
stopping effect at low energy. While for the net-charge moment
analysis, the case is opposite, as the total charged particle
multiplicity is larger for high energy than that of low energy. On
the other hand, with similar phase space coverage and number of
events, the moments analysis of net-charge distributions should have
larger errors than net-proton moments analysis at each collision
energy.

\begin{figure}[htb]
\begin{center}
\includegraphics[width=0.65\textwidth]{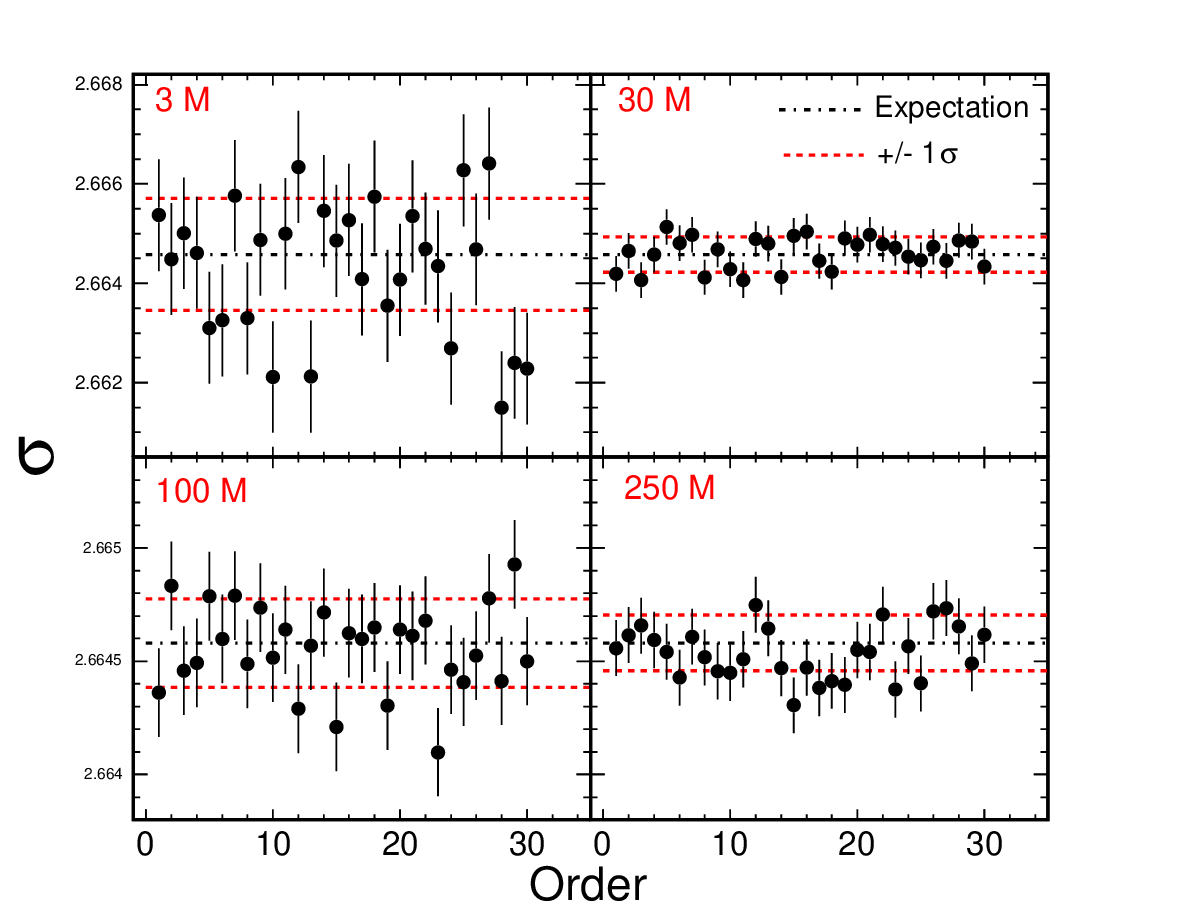}
\caption[]{Standard deviation ($\sigma$) of 30 samples that
independently and randomly generated from the Skellam distribution
with different number of events (3, 30, 100, 250 million). The
dashed lines are expectations and 1 $\sigma$ limits, respectively.}
\label{fig:sigma}
\end{center}\end{figure}

\begin{figure}[htb]
\begin{center}
\includegraphics[width=0.7\textwidth]{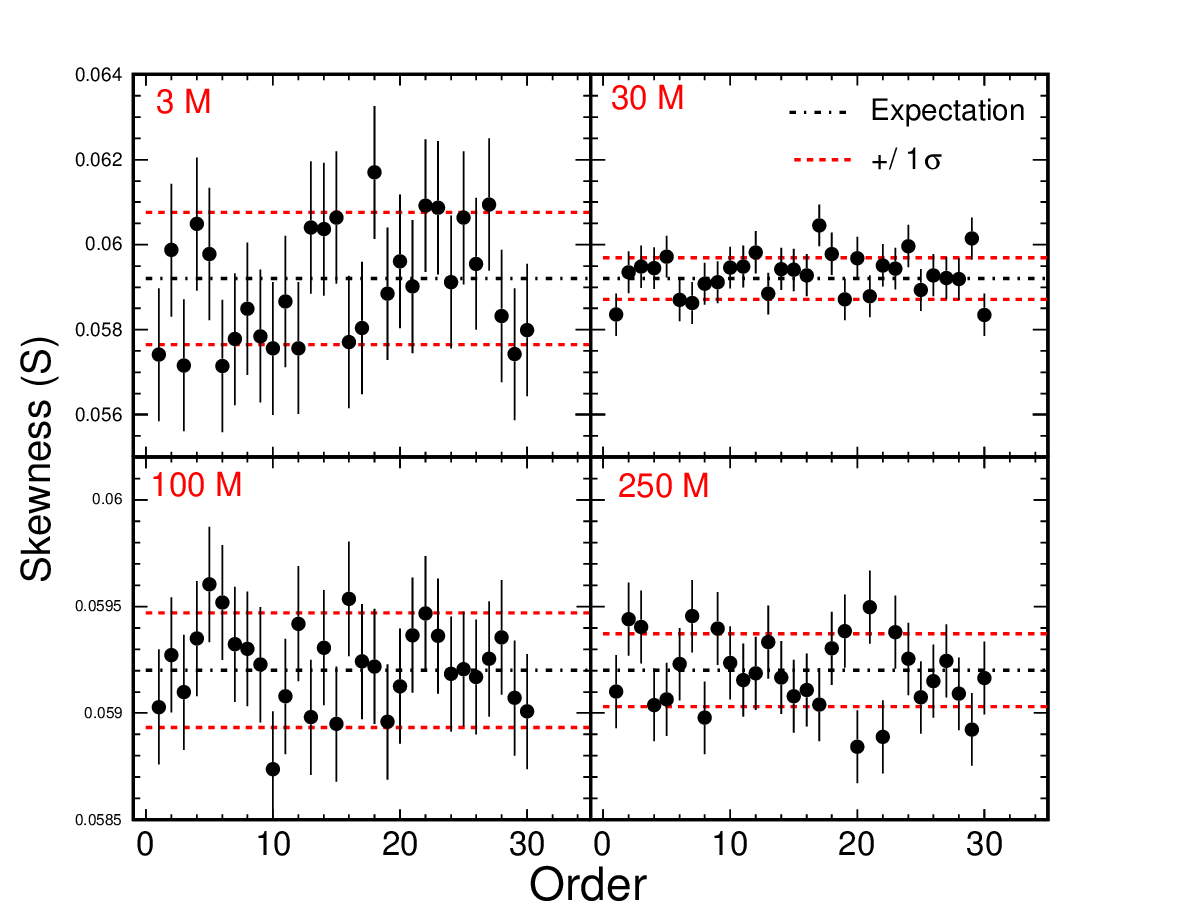}
\caption[]{Skewness ($S$) of 30 samples that independently and
randomly generated from the Skellam distribution with different
number of events (3, 30, 100, 250 million). The dashed lines are
expectations and 1 $\sigma$ limits, respectively. }
\label{fig:skewness}
\end{center}\end{figure}

\begin{figure}[htb]
\begin{center}
\includegraphics[width=0.7\textwidth]{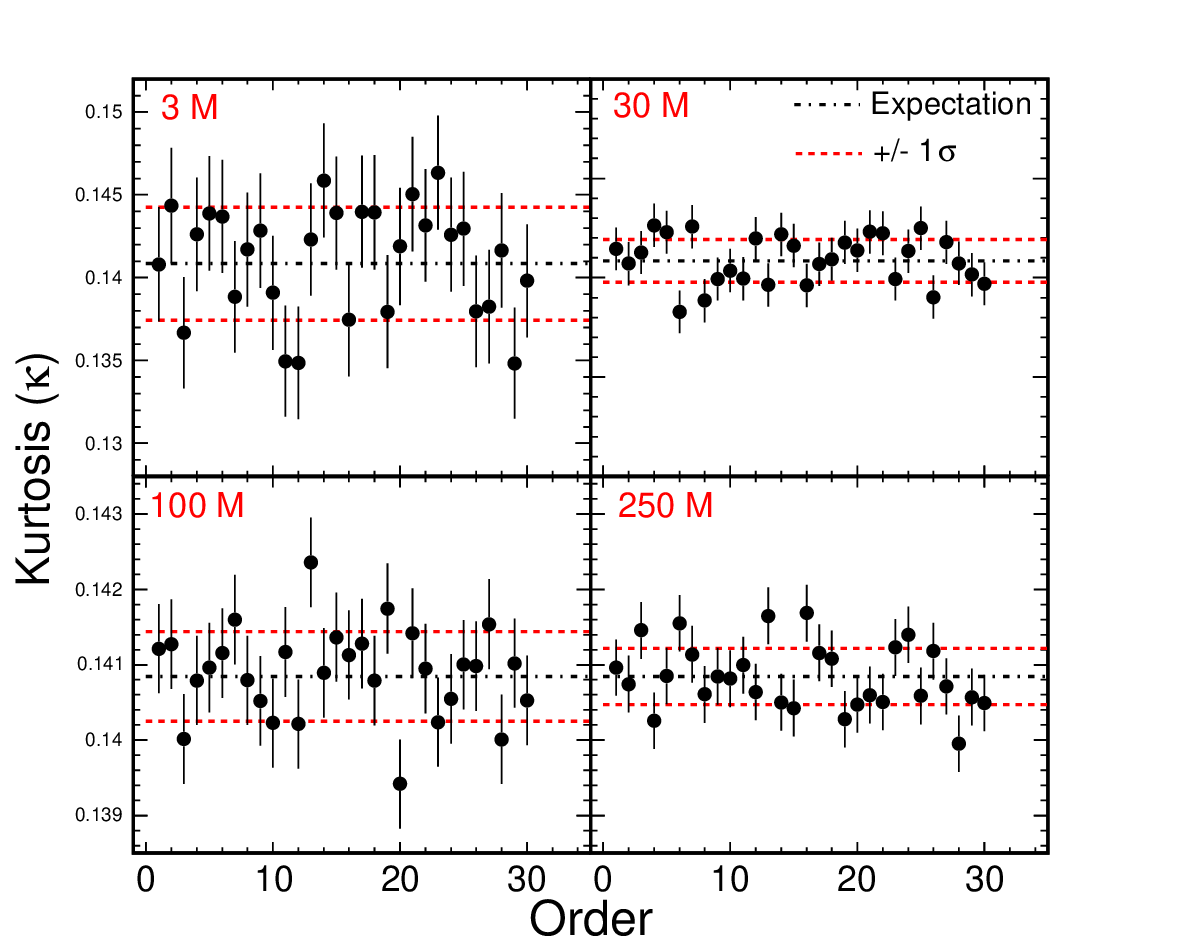}
\caption[]{Kurtosis ($\kappa$) of 30 samples that independently and
randomly generated from the Skellam distribution with different
number of events (3, 30, 100, 250 million). The dashed lines are
expectations and 1 $\sigma$ limits, respectively. }
\label{fig:kurtosis}
\end{center}\end{figure}

\begin{figure}[htb]
\begin{center}
\includegraphics[width=0.7\textwidth]{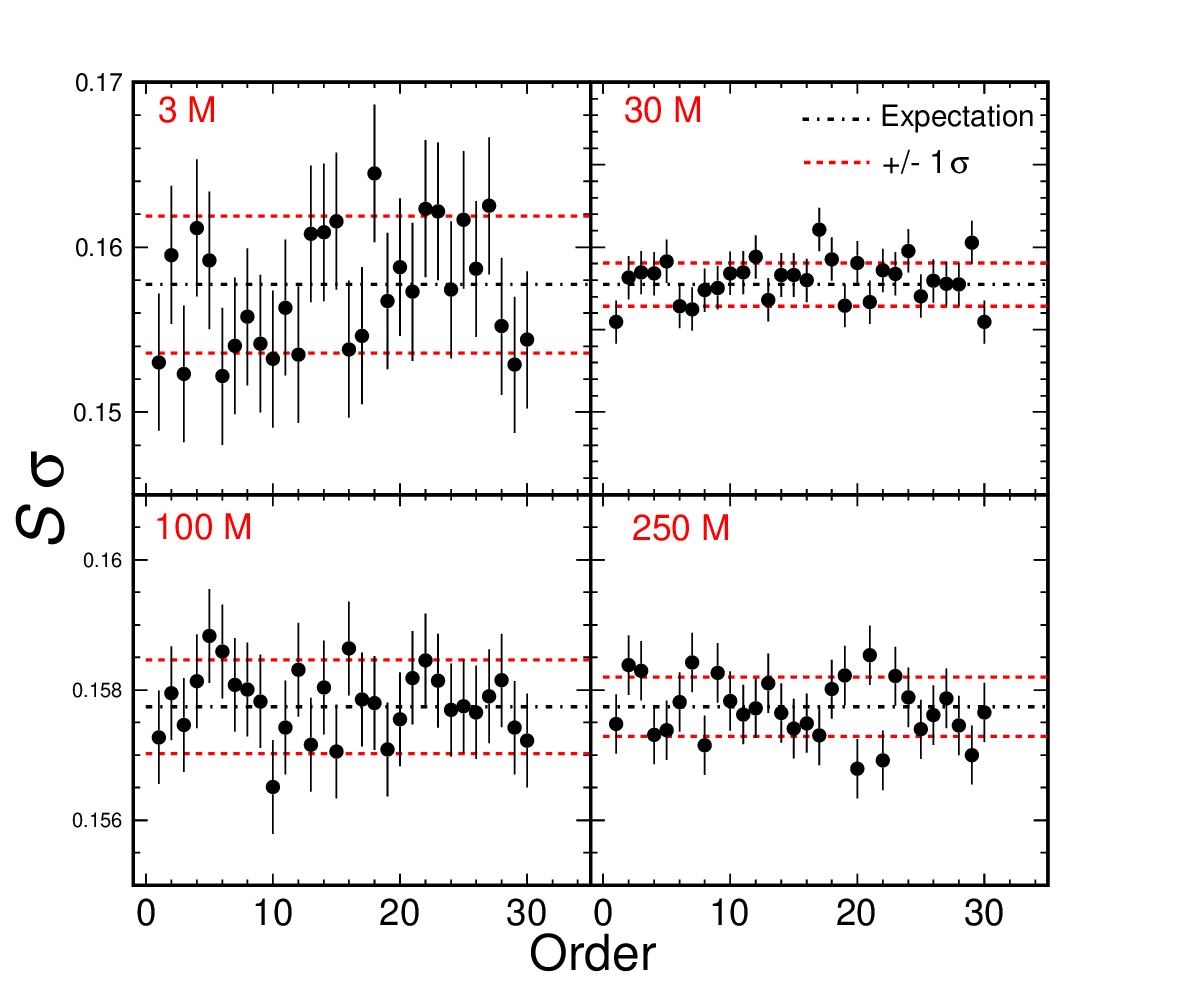}
\caption[]{Skewness ($S$) times standard deviation ($\sigma$) of 30
samples that independently and randomly generated from the Skellam
distribution with different number of events (3, 30, 100, 250
million). The dashed lines are expectations and 1 $\sigma$ limits,
respectively. } \label{fig:SD}
\end{center}\end{figure}

\begin{figure}[htb]
\begin{center}
\includegraphics[width=0.7\textwidth]{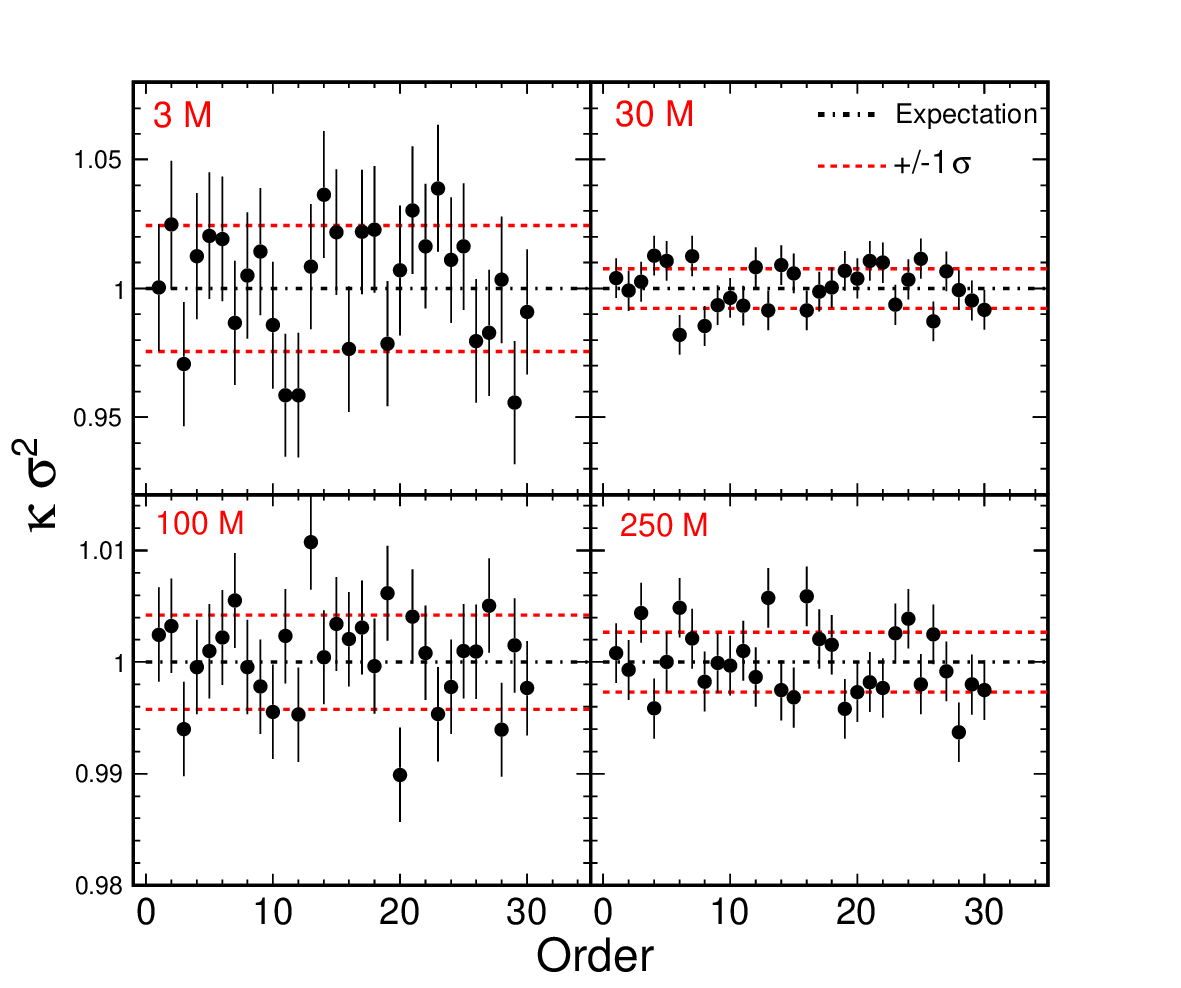}
\caption[]{Kurtosis ($\kappa$) times variance ($\sigma^2$) of 30
samples that independently and randomly generated from the Skellam
distribution with different number of events (3, 30, 100, 250
million). The dashed lines are expectations and 1 $\sigma$ limits,
respectively. } \label{fig:KV}
\end{center}\end{figure}

\begin{figure}[htb]
\begin{center}
\includegraphics[width=0.7\textwidth]{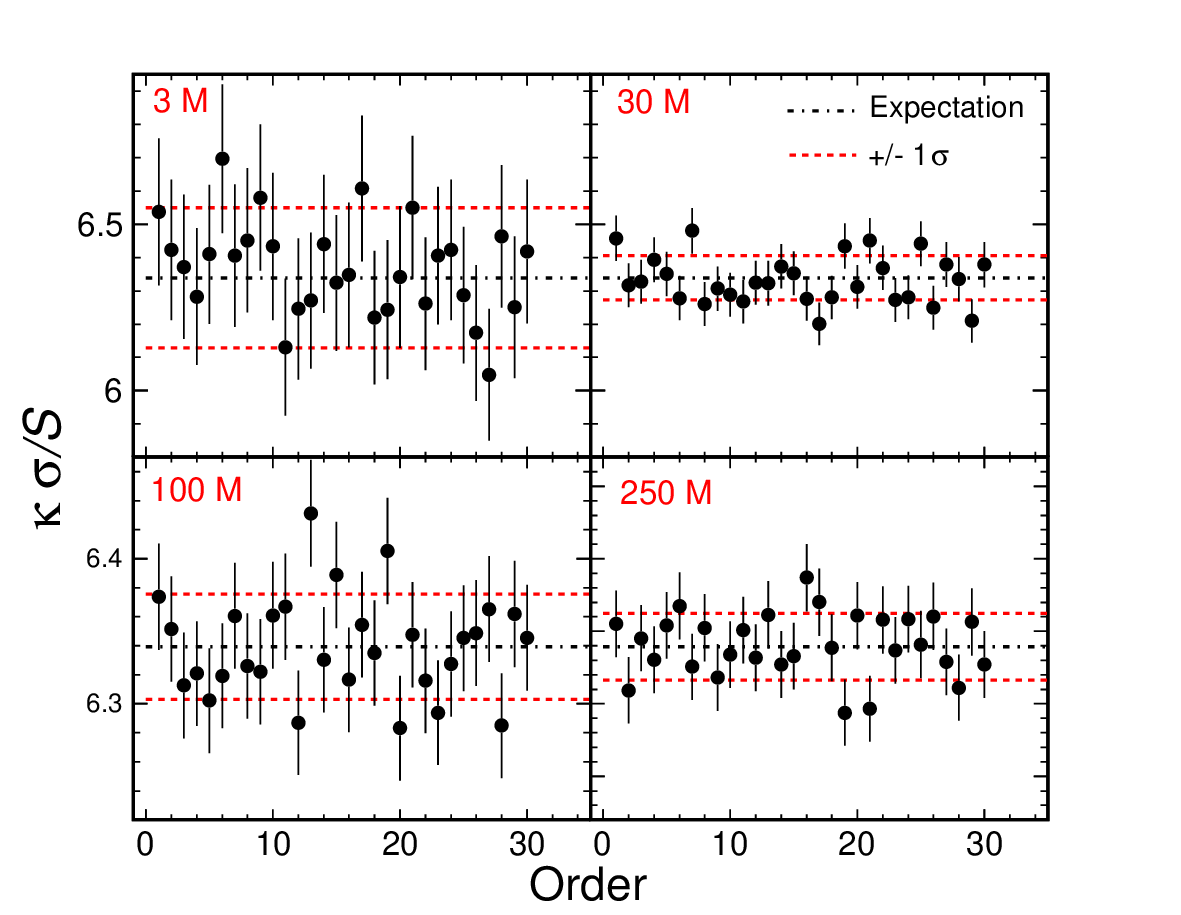}
\caption[]{ $\kappa \sigma/S$ of 30 samples that independently and
randomly generated from the Skellam distribution with different
number of events (3, 30, 100, 250 million). The dashed lines are
expectations and 1 $\sigma$ limits, respectively. } \label{fig:KDS}
\end{center}\end{figure}

\begin{figure}[htb]
\begin{center}
\includegraphics[width=0.7\textwidth]{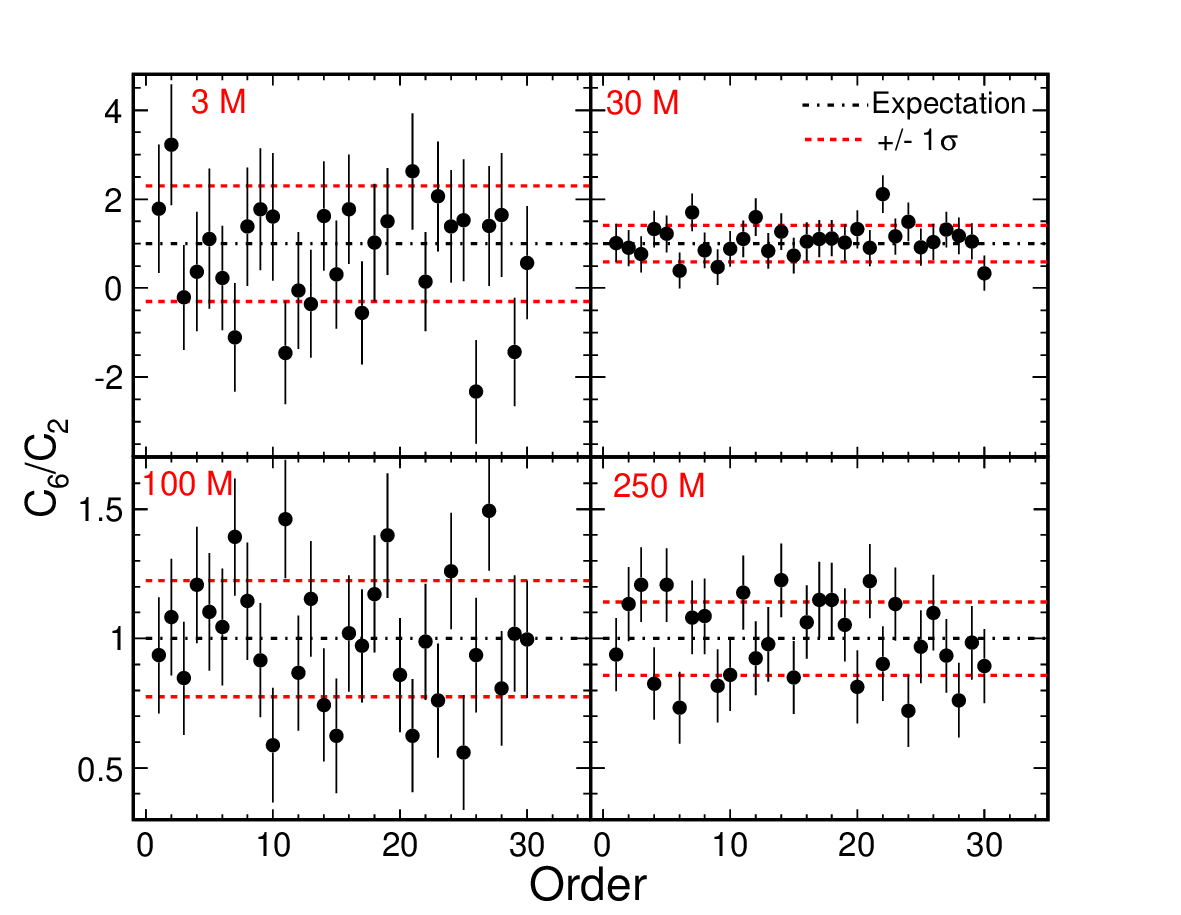}
\caption[]{Sixth to second order cumulant ratio ($C_{6}/C_{2}$) of
30 samples that independently and randomly generated from the
Skellam distribution with different number of events (3, 30, 100,
250 million). The dashed lines are expectations and 1 $\sigma$
limits, respectively. } \label{fig:c6c2}
\end{center}\end{figure}

\begin{figure}[htb]
\begin{center}
\includegraphics[width=0.7\textwidth]{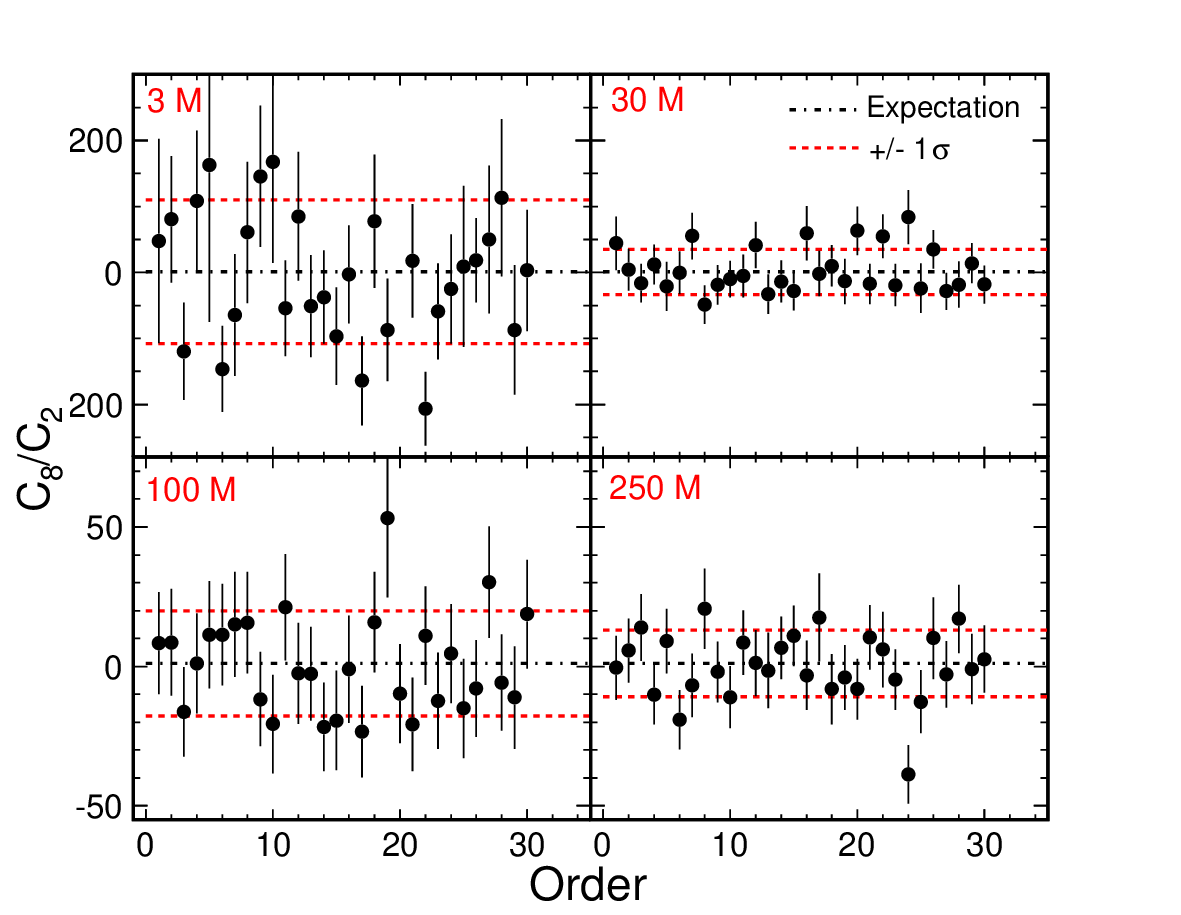}
\caption[]{Eighth to second order cumulant ratio ($C_{8}/C_{2}$) of
30 samples that independently and randomly generated from the
Skellam distribution with different number of events (3, 30, 100,
250 million). The dashed lines are expectations and 1 $\sigma$
limits, respectively.   } \label{fig:c8c2}
\end{center}\end{figure}

In the following, we will show various moments of 30 samples that
independently and randomly generated from the "Skellam" distribution
with different number of events (3, 30, 100 and 250 million) in Fig.
\ref{fig:sigma} to \ref{fig:c8c2}. The two parameters for "Skellam"
distribution is set to $\mu_1=4.11, \mu_2=2.99$. For comparison, we
also put the expected value for each moment and the one standard
deviation ($\sigma$) limit in those plots. For normal distribution,
the probability for the value staying within $\pm 1 \sigma$ around
expectation is about $68.3\%$, that means in each panel of Fig.
\ref{fig:sigma} to \ref{fig:c8c2} about 20 out of 30 points should
be in the range of $\pm 1 \sigma$. From Fig.\ref{fig:sigma} to
\ref{fig:c8c2}, we may find that all of the moments are well
satisfied this criteria and it indicates our error estimations for
those moments are reasonable and can reflect the statistical
properties of moments.

In Fig. \ref{fig:c8c2}, the eighth to second order cumulant ratio
has large error bars even with 250 million events. Reliable results
from this ratio need a large amount of statistics which may beyond
the number of events we have accumulated.

\section{Summary}
Higher moments of conserved quantities have been extensively studied
theoretically and experimentally. Due to the high sensitivity to the
correlation length and direct connection to the thermodynamic
susceptibilities, it can be used to probe the bulk properties, such
as chiral phase transition, critical fluctuation at critical point,
of hot dense matter created in the heavy ion collision experiment.
To perform precise higher moments measurement, the error analysis is
crucial for extracting physics message due to the statistic hungry
properties of the moments analysis. We have estimated the errors for
various moments that used in data analysis based on the Delta
theorem in statistics and probability theory. Monto Carlo simulation
is also done to check the validity of the error estimation. The
simulation shows that the error estimations for various moments can
reflect their statistical properties.

\section*{References}
\bibliography{MomentError}
\bibliographystyle{unsrt}
\newpage

\end{document}